\documentstyle[prd,aps]{revtex} 

\begin{document}
\title{Consistent discretization and loop quantum geometry}

\author{Rodolfo Gambini$^1$, and Jorge Pullin$^2$}
\address{1. Instituto de F\'{\i}sica, Facultad de Ciencias, 
Universidad
de la Rep\'ublica\\ Igu\'a esq. Mataojo, CP 11400 Montevideo, Uruguay}
\address{2. Department of Physics and Astronomy, 
Louisiana State University, Baton Rouge,
LA 70803-4001}

\date{September 14th 2004}
\maketitle
\begin{abstract}
  We apply the ``consistent discretization'' approach to general
  relativity leaving the spatial slices continuous. The resulting
  theory is free of the diffeomorphism and Hamiltonian constraints,
  but one can impose the diffeomorphism constraint to reduce its space
  of solutions and the constraint is preserved exactly under the
  discrete evolution. One ends up with a theory that has as physical
  space what is usually considered the kinematical space of loop
  quantum geometry, given by diffeomorphism invariant spin networks
  endowed with appropriate rigorously defined diffeomorphism invariant
  measures and inner products. The dynamics can be implemented as a
  unitary transformation and the problem of time explicitly solved or
  at least reduced to as a numerical problem. We exhibit the technique
  explicitly in $2+1$ dimensional gravity.
\end{abstract}

\bigskip

One of the central problems generated by the application of the rules
of quantum mechanics to general relativity is the problem of the
dynamics. When formulated canonically, general relativity has a
vanishing Hamiltonian, which has to be implemented as a constraint. In
the quantum geometry approach based on loop quantum gravity
\cite{review} the constraint has been implemented \cite{qsd}, but to
characterize the resulting theory in a way in which the dynamics of
general relativity is explicit remains a challenge.  We have recently
introduced a discrete approach to general relativity \cite{discrete,cosmo}
in which one approximates the continuum theory by a discrete theory
that is constraint free. This allows to make explicit progress in the
problem of the dynamics \cite{greece}.  One can formulate the quantum
theory in such a way that one chooses a physical variable as a clock
and describes the physics relationally in terms of conditional
probabilities \cite{deco}.  The approach however appears radically
different from usual loop quantum gravity, and does not seem to
incorporate due to the discreteness, many of the attractive
mathematical structures that have been developed in loop quantum
gravity. In particular the characterization of states in terms of knot
invariants and the existence of a rigorous mathematical arena to
describe the theory.

In this paper we would like to bridge the gap between these two
approaches. We will analyze the consequences of applying our
consistent discretization technique to general relativity in the
time-like direction, while keeping the spatial slices continuous. The
resulting canonical theory has discrete time evolution and is free of
constraints, as is usually the case in consistent discretizations. One
can however further restrict the dynamics of the theory by imposing
the diffeomorphism constraint of the usual continuum theory.
Remarkably, the generator of diffeomorphisms of the continuum theory
is preserved by the discrete evolution. This, in fact, is the key
observation of this paper. If one starts from a formulation of general
relativity based on Ashtekar's variables and performs this
construction, one would have a theory that could be quantized using
the usual tools of loop quantum gravity. In particular the states will
be functions of spin networks that are annihilated by the
diffeomorphism constraint and one can introduce the
Ashtekar-Isham-Lewandowski measure and theory of integration. The
usual well defined quantum operators like the area and the volume will
exist and be well defined, except that in the discrete theory the
total volume of the slice will be an observable since there are no
further constraints.  The resulting theory could be used as a basis to
construct a relational quantization a la Page--Wootters \cite{PaWo}
and introduce a physical clock  that defines evolution
through relational probabilities. One therefore has a mathematically
well defined arena in which to complete the quantization of general
relativity through a well defined procedure that can be carried out
entirely, the only challenge left for completing the construction of
the quantum theory being of computational nature. Of course, there is
still the issue of if the resulting theory will have a correct
semiclassical limit.

It may be argued  that discretizing the temporal
evolution while keeping space continuous is unnatural. After all space
and time are supposed to be treated in the same footing in general
relativity.  However, it should be noted that although one starts from
a spatially continuous classical theory, the loop quantization
naturally introduces a discrete structure in space. Therefore the
final theory will end up with both space and time discrete.

Let us illustrate the idea in the case of a $1+1$-dimensional theory
in the continuum with an action $S=\int dt dx L({q}(x),
\dot{q}(x),q'(x),q''(x))$ where to simplify notation we are
considering only one variable and use the primes to denote derivatives
with respect to the spatial coordinate. One then discretizes time and the
action becomes $S =\sum_n L(n,n+1)$ where $L(n,n+1)$ is obtained from
the Lagrangian replacing the time derivatives by
$\dot{q}=(q_{n+1}-q_n)/\epsilon$. We assume the action has the form of
a first order theory with constraints,
\begin{equation}
  L(n,n+1)=\int dx \pi_n(x) (q_{n+1}(x)-q_n(x)) -
\epsilon \int dx H(q_n(x),\pi_n(x)) -\int dx N_n(x) \phi(q_{n}(x),\pi_n(x))
\end{equation}
where we have assumed that the theory may have a Hamiltonian $H$ (in
the case of general relativity $H$ vanishes and the theory loses all
information about the discretization step $\epsilon$) and
constraint(s) $\phi(q_n(x),\pi_n(x))$. To simplify notation we are not
making explicit the dependence of the Hamiltonian and the constraints
on spatial derivatives of the fields, but this is allowed in our
approach.  One now introduces the canonically conjugate variables,
\begin{eqnarray}
  P_{n+1}(y) &\equiv& {\delta L(n,n+1) \over \delta q_{n+1}(y)} = \pi_n(y),
\qquad   
P^\pi_{n+1}(y) \equiv {\delta L(n,n+1) \over \delta \pi_{n+1}(y)} = 0,
\qquad
P^N_{n+1}(y) \equiv {\delta L(n,n+1) \over \delta N_{n+1}(y)} =0,
\\
\label{3}
  P_n(y) &\equiv& -{\delta L(n,n+1) \over \delta q_{n}(y)} = \pi_n(y)
-{\delta \over \delta q_n(y)} \left[
\epsilon \int dx H(q_n(x),\pi_n(x)) +\int dx N_n(x) \phi(q_{n}(x),\pi_n(x))
\right],
\\
  P^\pi_n(y) &\equiv& -{\delta L(n,n+1) \over \delta \pi_{n}(y)} =
-(q_{n+1}(y)-q_n(y)) 
-{\delta \over \delta \pi_n(y)} \left[
\epsilon \int dx H(q_n(x),\pi_n(x)) +\int dx N_n(x) \phi(q_{n}(x),\pi_n(x))
\right],
\\
  P^N_{n}(y) &\equiv& {\delta L(n,n+1) \over \delta N_{n}(y)}
  =\phi(q_{n}(y),\pi_n(y)). 
\label{7}
  \end{eqnarray}

One now can eliminate the variable $\pi$ and its canonical momentum
$P^\pi$ and end up with a theory entirely given in terms of $q$ and
its canonically conjugate momentum $P$. The theory is constraint
free since equation (\ref{7}) now becomes
$\phi(q_n(x),P_{n+1}(x))=0$ and therefore it is not a constraint, in
the sense that it does not constrain variables at the same time level.
If one now substitutes $P_{n+1}(x)$ making use of (\ref{3}) one
gets a differential equation that determines the Lagrange
multiplier $N_n(x)$. That is, we have a theory that is constraint
free at the expense of determining the Lagrange multipliers, as 
is usually the case in the consistent discretization approach. 
The resulting theory has more degrees of freedom than the continuum
theory it attempts to approximate. This is due to the fact that a 
single solution of the continuum theory can be approximated by 
several, different, solutions to the discrete theory. 
We will now turn our attention to the specific case of general
relativity and proceed to reduce the extra number of degrees of
freedom by choosing a sector of solutions of the discrete theory
that is preserved upon evolution. The sector is chosen by requiring
that the usual diffeomorphism constraint of general relativity be
satisfied. It might appear surprising at first that the requirement that
the constraint be satisfied is preserved by the discrete evolution.

In order to see this, let us consider the action for general relativity
written in terms of Ashtekar's variables \cite{asle},
\begin{equation}
  S=\int dt d^3x \left(\tilde{P}^a_i F_{0a}^i -N^a C_a -N C\right) 
\end{equation}
where $N,N^a$ are Lagrange multipliers, $\tilde{P}^a_i$ are densitized
triads, and the diffeomorphism and
Hamiltonian constraints are given by,
\begin{eqnarray}
  C^a&=&\tilde{P}^a_i F_{ab}^i\\
  C &=&{\tilde{P}^a_i \tilde{P}^b_j \over \sqrt{\rm det} q}
\left(\epsilon^{ijk} F_{ab}^i -
(1+\beta^2) K_{[a}^i K_{b]}^j \right)
\end{eqnarray}
where $\beta K_a^i\equiv \Gamma_a^i-A_a^i $ and $\Gamma_a^i$ is the
spin connection compatible with the triad, and $q$ is the three
metric. We now proceed to discretize time.  The action now reads,
\begin{eqnarray}
  S&=&\int dt d^3x \left[{\rm Tr}\left( \tilde{P}^a
\left(A_a(x) -V(x) A_{n+1,a}(x) V^{-1}(x)+\partial_a(V(x)) V^{-1}(x)\right)
\right)\right.\\
&&\left.-N^a C_a -N C+ \mu \sqrt{\rm det} q  {\rm Tr}\left(V(x)V^\dagger(x)-1\right)\right] \nonumber
\end{eqnarray}
In the above expression $V(x)=V_I T^I$ is the parallel transport
matrix along a time-like direction and $F_{0a}$ is approximated by the
holonomy along a plaquette that is finite in the ``time-like''
direction and infinitesimal in the ``space-like'' direction and
$T^0=1/\sqrt{2}$ and $T^a=-i\sigma^a/\sqrt{2},a=1..3$ and $\sigma$'s
are the Pauli matrices and the coefficients $V_I$ are real. We have
omitted the subscript $n$ to simplify the notation and kept it in the
quantities that are evaluated at $n+1$. The last term involves a
Lagrange multiplier $\mu$ and is present in order to enforce the fact
that the parallel transport matrices are unitary.  We notice that the
$SU(2)$ gauge invariance is preserved in the semi-discrete theory.
This in turn implies that Gauss' law for the momentum canonically
conjugate to the connection, $\tilde{E}^a_{n+1} \equiv V^{-1}
\tilde{P}^a V$ is preserved automatically upon evolution. We do not
work this out explicitly here for reasons of space, the reader can
refer to the example of BF theory we present later in this paper to
see how the conservation works in detail, the mechanism is similar to
the one in general relativity.

We now consider a spatial (time independent) infinitesimal
transformation ${x'}^a=x^a+v^a(x)$. It is immediate to see that the
action is invariant. All variables transform as they do in the
continuum action, and $V(x)$ transforms as a scalar. The only question
could be the first term, but since the transformation is time
independent the terms at $n+1$ and $n$ transform appropriately.
Applying Noether's theorem, there is a resulting conserved quantity
that can be readily computed using the Lagrange equations, and the
resulting quantity is $C_a=\tilde{E}^b_i F_{ab}^i$ that is, the
diffeomorphism constraint of the continuum theory. We have checked the
conservation explicitly.

Let us outline how would one complete the quantization. The central
element is to implement the canonical transformation that evolves the
variables from $n$ to $n+1$ as a unitary operator.  
Quantum states will be functions of the connection $\Psi[A]$ that are
invariant under diffeomorphisms and gauge transformations.  For
example one could consider cylindrical functions based on spin
networks.  We need to construct the unitary operator $\Psi_{n+1}[A']=
\int DA U(A',A) \Psi_n[A]$. Since the canonical transformation that
implements the evolution is generated by the Lagrangian, the unitary
operator in the configuration basis is given by the exponential of the
Lagrangian \cite{Dirac,cosmo} viewed as function of $A_n$ and
$A_{n+1}$. In practice to compute the Lagrangian as a function of
these variables one needs to solve the equations of motion between $n$
and $n+1$. In situations of interest this could be achieved
numerically, for instance, or through other approximation schemes. To
make the calculation feasible numerically one will have to choose to
work in a subspace of states to make computations finite.  The
calculations in situations of great generality will be hard, but the
point to emphasize here is that there is no conceptual obstacle to
carrying them out. In other words, what we have here is a concrete
proposal for doing numerical quantum gravity.

A point to be noted is that the calculation of the unitary evolution
operator can be carried out in a context that is not diffeomorphism
invariant. Given that the canonical transformation preserves the
diffeomorphism generator, one has for the unitary operator that
$V_\phi U(A',A) V^\dagger_\phi = U(\phi A',\phi A)$ where
$V_\phi$ is the generator of a finite diffeomorphism $\phi$. If one
starts from wavefunctions that are invariant under diffeomorphisms,
i.e., $V_\phi \Psi[A] = \Psi[A]$, then one has that $U(A',A)$ has to
satisfy,
\begin{eqnarray}
  \Psi_{n+1}[A'] &=&\int DA U[A',A] \Psi_n[A] = \int DA 
  U[A',A]V_\phi \Psi_n[A]
  =\int DAU[A',\phi^{-1} A] \Psi_n[A]\nonumber\\
&=&\int DAU[\phi A',A]\Psi_n[A]= \Psi_{n+1}[\phi A'],
\end{eqnarray}
and the integrals can be rigorously defined using the
Ashtekar--Lewandowski integration theory developed on cylindrical
functions and their Cauchy completions \cite{review}.

Therefore the evolution yields a diffeomorphism invariant state. A
similar comment applies to the invariance under Gauss' law (gauge
invariance).

Once one has the explicit evolution of the wavefunctions, then one can
choose a physical time from among the variables of the problem and
construct a relational quantum theory as outlined in \cite{greece,deco}.

It is clear that carrying out the proposal in detail in situations of
interest with local degrees of freedom will require significant
computational effort, even in simplified examples like the Gowdy
cosmologies. To present a concrete illustration of the technique in a
non-trivial setting that allows to implement things in detail we will
discuss $2+1$ dimensional $SU(2)$ BF theory and see that the approach
yields the correct expected results. This is of some interest since this
theory is equivalent to Euclidean general relativity in $2+1$ dimensions.

We start with the standard action for BF theory $S=\int d^3x {\rm
  Tr}\left(B\wedge F\right)$ and we discretize the ``temporal'' direction
(we label the ``spatial'' directions $1,2$ and the temporal one $0$), 
$S=\sum_n \int d^2x L(n,n+1)$ with,
\begin{eqnarray}
L(n,n+1)&=& {\rm Tr}\left\{B_0(x) F_{12}(x)+
 B_1(x)\left(A_2(x) -V(x) A_{n+1,2}(x) V^{-1}(x)+\partial_2 (V(x)) V^{-1}(x)\right)
\right.\nonumber\\
&&\left.+
 B_2(x)\left(V(x)A_{n+1,1}(x)V^{-1}(x)+ 
V(x) \partial_1 V^{-1}(x)-A_1(x)\right)+\mu \left(V(x)V^\dagger(x)-1\right)\right\}
\end{eqnarray}
and we are using the same notation as in the general relativity case.
In two of the three terms in the action we have approximated the
curvature by a holonomy along a plaquette that is finite in the
``time-like'' direction and infinitesimal in the ``space-like''
direction. As before, the $SU(2)$ gauge invariance is preserved in the
semi-discrete theory.

We now build the canonical theory as is usually done in consistent
discretizations by defining the canonical conjugate momenta. All of
them vanish except the conjugates to the components $1,2$ of the
connection, which we suggestively call $E^{1,2}$ and are given by,
\begin{equation}
  E^i(x)_{n+1}=V^{-1}(x)B^iV(x),\quad i=1,2,
\end{equation}
and $B^i=\epsilon^{ij} B_j$.

The definition of the canonical momenta to the $B^i_n$'s yields
evolution equations for the $A_i$'s,
\begin{eqnarray}
  P^{B^1}&=&0=A_2(x)-V A_{n+1,2}(x) V^{-1}(x)+\partial_2 (V(x))V(x)^{-1},\\
  P^{B^2}&=&0=-A_1(x)+V A_{n+1,1}(x) V^{-1}(x)-\partial_1 (V(x))V(x)^{-1}.
\end{eqnarray}
The momentum of $B^0$ vanishes and yields as equation the constraint
$F_{12}=0$ and the momentum of $V$, called $P^V_{n+1}$ also vanishes
and this yields Gauss' law $P^V_n V_n\equiv D_{n+1,a} E^a_{n+1}=0$.
The constraint implies that the connection is pure gauge and the
evolution equations for the connection guarantee that if one starts
from a connection that is pure gauge it evolves into a pure gauge
connection. We are omitting other evolution equations (like the ones
for the momenta) for reasons of space.

We can outline the quantization. We choose a connection representation
in which $A$ is multiplicative and $E$ is a functional derivative and
they have canonical commutation relations. The evolution equations
become operatorial equations among operators in a Heisenberg-like
representation where the role of the ordinary time would be played by
the variable $n$.  If one wishes to construct the unitary
transformation that would implement the dynamics in the quantum
theory, one proceeds in the following way.  One computes the
expectation value of the equations of motion between eigenstates of
the operator $A$ at instants $n+1$ and $n$. This allows to infer, by
solving the resulting functional equations that the value of
\begin{eqnarray}
U(A',A)&=&<A',n+1|A,n>\nonumber\\
&&=\delta\left(A'_1-V^{-1}A_1 V+\partial_1 (V^{-1}) V\right)
\delta\left(A'_2-V^{-1}A_2 V+\partial_2 (V^{-1}) V\right) 
\exp\left({\rm Tr}\int B_0 F\right),\label{unitary}
\end{eqnarray}
and this expression satisfies equation (10).

Since one has constraints one needs to impose them on the space of
states, and this corresponds to the usual space of states of BF
theory as discussed by Ooguri and more recently by Noui and Perez 
\cite{Ooguri}. As usual for BF
theories, the constraints imply that the theory is spatially
diffeomorphism invariant. Notice that the unitary transformation
(\ref{unitary}) depends on two free functions, $V$ and $B_0$.
However, on the physical space these functions can be freely
chosen without affecting the evolution, as can be seen by
inspection of (\ref{unitary}).

Summarizing, we have observed that one can introduce the ``consistent
discretization'' technique in general relativity and other constrained
theories in which one keeps spatial slices continuous and discretizes
time. The resulting semi-discrete theory has constraints that can be
solved by going to the loop representation, and therefore can be
explicitly handled without conceptual problems but one can
consistently impose on its space of states the diffeomorphism
constraint as a further restriction. One ends up with a theory that
has as physical space the usual diffeomorphism invariant kinematical
structure of loop quantum gravity and one can take advantage of the
various mathematical developments and physical results of that arena.

The idea that we consider continuous space and
discrete time within the consistent approach was suggested to us by
Lee Smolin.  This work was supported in part by grants NSF-PHY0244335,
and by funds of the Horace C. Hearne Jr. Institute for Theoretical
Physics.

\end{document}